\documentclass[%
 reprint,
 amsmath,amssymb,
 aps,
]{revtex4-2}

\usepackage{graphicx}

\makeatletter
\newcommand*\bigcdot{\mathpalette\bigcdot@{.5}}
\newcommand*\bigcdot@[2]{\mathbin{\vcenter{\hbox{\scalebox{#2}{$\m@th#1\bullet$}}}}}
\makeatother

\newcommand{\deq}{\stackrel{\bigcdot}{=}}

\usepackage{color}
\usepackage{tensor}
\usepackage[all,cmtip]{xy}
\usepackage{dcolumn}
\usepackage{bm}

\newcommand{\overbar}[1]{\mkern 1.5mu\overline{\mkern-1.5mu#1\mkern-1.5mu}\mkern 1.5mu}

\def\TT{{T\overbar{T}}}

\newcommand{\Tr}[1]{\text{Tr}\left[{#1}\right]}

\newcommand{\fL}{\mathfrak{L}}
\newcommand{\fJ}{\mathfrak{J}}

\begin{document}


\title{\boldmath Infinite Family of Integrable Sigma Models Using Auxiliary Fields}

\author{Christian Ferko}\email{caferko@ucdavis.edu}
\affiliation{%
Center for Quantum Mathematics and Physics (QMAP),
Department of Physics \& Astronomy, University of California, Davis, CA 95616, USA.}

\author{Liam Smith}\email{liam.smith1@uq.net.au}
\affiliation{%
School of Mathematics and Physics, University of Queensland,
St Lucia, Brisbane, Queensland 4072, Australia.}%

\date{\today}

\begin{abstract}

We introduce a class of $2d$ sigma models which are parameterized by a function of one variable. In addition to the physical field $g$, these models include an auxiliary field $v_\alpha$ which mediates interactions in a prescribed way. We prove that every theory in this family is classically integrable, in that it possesses an infinite set of conserved charges in involution, which can be constructed from a Lax representation for the equations of motion. This class includes the principal chiral model (PCM) and all deformations of the PCM by functions of the energy-momentum tensor.

\end{abstract}

\maketitle

\section{Introduction}

Integrable models in two dimensions offer rare examples of interacting field theories which can nonetheless be analyzed exactly, both at the classical and quantum levels. The dynamics of such models are tightly constrained by the presence of hidden symmetries, which can sometimes be leveraged to ``solve'' the theories. It is therefore useful to identify many examples of such integrable models, which has been a topic of great recent interest. These integrable $2d$ theories find applications in many areas of physics, such as in studies of classical and quantum strings at finite tension~\cite{Arutyunov:2009ga,Beisert:2010jr,Demulder:2023bux} and in condensed matter descriptions of spin chains (see e.g.~\cite{HALDANE1983464,Faddeev:1996iy,Essler_Frahm_Göhmann_Klümper_Korepin_2005}).

A fruitful method for generating new integrable theories is to deform existing ones in a way which preserves integrability. One popular starting point for this enterprise is the principal chiral model (PCM) \cite{PhysRevLett.30.1343,PhysRevLett.30.1346}, which is an integrable $2d$ model that shares certain properties with $4d$ Yang-Mills theory \cite{POLYAKOV197579,Polyakov:1975yp,Migdal:1975zg,PhysRevLett.69.873}.
The PCM admits many integrable deformations, some of which are reviewed in \cite{Zarembo:2017muf,Orlando:2019his,Seibold:2020ouf,Klimcik:2021bjy,Hoare:2021dix}; for instance, one can add a Wess-Zumino term \cite{WESS197195,Novikov:1982ei,Witten:1983ar,ABDALLA1982181}, perform a Yang-Baxter deformation \cite{Klimcik:2002zj,Klimcik:2008eq}, implement a $\lambda$-deformation \cite{Sfetsos:2013wia}, and so on. 

In this letter, we introduce a new, infinite class of integrable sigma models by deforming the PCM via the activation of specific interactions that preserve integrability. The key technical ingredient in our construction is the coupling of the physical field in the PCM to an auxiliary field with algebraic equations of motion. There are many examples in field theory where the introduction of auxiliary fields can impose a desirable constraint, such as in the Pasti-Sorokin-Tonin formalism \cite{Pasti:1995tn,Pasti:1996vs,Pasti:1997gx} which uses an auxiliary vector to implement a chirality condition on a tensor field in $6d$, or the Ivanov-Zupnik (IZ) auxiliary field formalism \cite{Ivanov:2002ab,Ivanov:2003uj} which imposes electric-magnetic duality invariance in theories of $4d$ electrodynamics. In our case, the inclusion of an auxiliary field imposes that the model remain integrable in a particularly simple way, while nonetheless allowing for quite general interactions.

An important part of the analysis of a classically integrable field theory is the construction of an infinite set of Poisson-commuting conserved charges. For the PCM and many of its deformations, this procedure can be performed in two steps. The first step is to establish that the equations of motion for the model can be encoded in the flatness of a one-form called the Lax connection. The second step is to compute a particular Poisson bracket involving the Lax connection and demonstrate that it takes a special form considered by Maillet \cite{MAILLET198654,MAILLET1986401}. This Maillet form of the Poisson bracket is parameterized by an $r$-matrix $r_{12} ( z, z' )$, which for many cases can be written in a standard way in terms of a ``twist function'' $\varphi ( z )$. Integrable deformations of the PCM often modify the form of this twist function $\varphi ( z )$; see \cite{Lacroix:2018njs} for examples.

In this work, we perform these two steps for our new family of models. We find that, unlike some other integrable deformations of the PCM, the twist function is not modified for theories within our class, although the dependence of the Lax connection on the fields is changed. This behavior is similar to that of homogeneous Yang-Baxter deformations \cite{Kawaguchi:2014qwa,vanTongeren:2015soa,Borsato:2021fuy}, which can be interpreted via twisted boundary conditions (it would be interesting to explore whether a similar interpretation applies to our models). As we will see, the proof of classical integrability for our models is nearly as simple as that for the PCM, even though a generic model in this family exhibits highly non-linear interactions. This illustrates the broader theme that one can often introduce interactions while preserving a desirable structure of a field theory, such as integrability, through auxiliary field techniques.

The structure of this letter is as follows. In Section \ref{sec:models}, we introduce the family of sigma models of interest in this work and derive the equations of motion for these theories. In Section \ref{sec:weak}, we prove that the equations of motion for these models are equivalent to flatness of a Lax connection. Section \ref{sec:strong} computes the Poisson bracket of the spatial component of this Lax connection and shows that it takes the Maillet form, which establishes the existence of an infinite set of conserved charges in involution. In Section \ref{sec:stress}, we show that our family of models includes all deformations of the PCM by functions of the stress tensor, such as $\TT$ and root-$\TT$. Finally, Section \ref{sec:conclusion} summarizes and presents directions for future research.

\section{Auxiliary Field Sigma Models}\label{sec:models}

We now describe the class of models which are of interest in the present work. The physical degree of freedom is a group-valued field $g : \Sigma \to G$ which maps a flat two-dimensional spacetime
$\Sigma$ with coordinates $\sigma^\alpha = ( \tau, \sigma )$ into a Lie group $G$. We write $\mathfrak{g}$ to denote the Lie algebra of $G$. The left-invariant Maurer-Cartan form is defined by $j = g^{-1} d g$, and the pullback of this form to $\Sigma$ is
\begin{align}
    j_\alpha = g^{-1} \partial_\alpha g \, .
\end{align}
It will be convenient to use light-cone coordinates on $\Sigma$,
\begin{align}
    \sigma^{\pm} = \frac{1}{2} \left( \tau \pm \sigma \right) \, ,
\end{align}
which are lowered or raised with the metric $\eta_{+ - } = \eta_{- +} = - 2$ or inverse metric $\eta^{+ -} = \eta^{- +} = - \frac{1}{2}$. The field $j_\alpha$ satisfies the Maurer-Cartan identity, which is written in light-cone coordinates as
\begin{align}\label{maurer_cartan}
    \partial_+ j_- - \partial_- j_+ + [ j_+ , j_- ] = 0 \, .
\end{align}
In our notation, the Lagrangian of the PCM is written
\begin{align}\label{PCM_lagrangian}
    \mathcal{L}_{\text{PCM}} = \frac{1}{2} \eta^{\alpha \beta} \mathrm{tr} ( j_\alpha j_\beta ) = - \frac{1}{2} \mathrm{tr} ( j_+ j_- ) \, .
\end{align}
We now introduce an additional Lie algebra valued field $v_\alpha$, which is not a physical degree of freedom, but merely an auxiliary field with algebraic equations of motion. One can take traces to build scalars from $v_\alpha$, such as
\begin{align}
    \nu = \mathrm{tr} ( v_+ v_+ ) \mathrm{tr} ( v_- v_- ) \, .
\end{align}
The family of models which we wish to study are described by Lagrangians of the form 
\begin{align}\label{integrable_family_auxiliaries}
    \mathcal{L} &= \frac{1}{2} \mathrm{tr} ( j_+ j_- ) + \mathrm{tr} ( v_+ v_- ) + \mathrm{tr} ( j_+ v_- + j_- v_+ ) + E ( \nu ) \, ,
\end{align}
which are parameterized by an arbitrary interaction function $E ( \nu )$ that depends on the single variable $\nu$. The structure of this family of Lagrangians is inspired by the Ivanov-Zupnik formulation of theories of duality-invariant electrodynamics in four dimensions \cite{Ivanov:2002ab,Ivanov:2003uj}, which also involves an interaction function that depends on a single variable constructed from auxiliary fields. 

Let us first consider the equations of motion associated with the Lagrangian (\ref{integrable_family_auxiliaries}). Varying the auxiliary field $v_\alpha$ gives rise to the Euler-Lagrange equation
\begin{align}\label{aux_eom}
    j_{\pm} = - v_{\pm} - 2 E' v_{\mp} \mathrm{tr} ( v_{\pm} v_{\pm} ) \, ,
\end{align}
where we write $E' = \frac{d E}{d \nu}$ for the derivative of $E$.

On the other hand, varying the group-valued field $g$ as $\delta g = g \epsilon$, under which the Maurer-Cartan form varies as $\delta j_{\pm} = \partial_{\pm} \epsilon + [ j_{\pm} , \epsilon ]$, gives the $g$-field equation of motion
\begin{align}\label{j_eom}
    \partial_+ j_- + \partial_- j_+ = 2 \left( [ v_- , j_+ ] + [ v_+, j_- ] - \partial_+ v_- - \partial_- v_+ \right) \, .
\end{align}
The interpretations of the two Euler-Lagrange equations (\ref{aux_eom}) and (\ref{j_eom}) are rather different. While (\ref{j_eom}) is a true dynamical condition, which implies that the physical field $g$ is on-shell, equation (\ref{aux_eom}) is merely a constraint imposed by the auxiliary field. We introduce the symbol $\deq$ to denote equality between two quantities which holds when equation (\ref{aux_eom}) is satisfied. For instance, (\ref{aux_eom}) implies
\begin{align}
    [ v_- , j_+ ] + [ v_+, j_- ] \deq 0 \, ,
\end{align}
so the $g$-field equation of motion (\ref{j_eom}) can be written as
\begin{align}
    \partial_+ \left( j_- + 2 v_- \right) + \partial_- \left( j_+ + 2 v_+ \right) \deq 0 \, ,
\end{align}
when the auxiliary field equation of motion (\ref{aux_eom}) is satisfied. It is convenient to define the new quantity
\begin{align}
    \mathfrak{J}_{\pm} = - \left( j_{\pm} + 2 v_{\pm} \right) \, ,
\end{align}
which allows us to express the Euler-Lagrange equation for $g$ as a conservation equation $\partial_\alpha \mathfrak{J}^\alpha \deq 0$. Here $j_\alpha$ is flat but not conserved, while $\fJ_\alpha$ is conserved but not flat.

Note that, when the interaction function $E ( \nu )$ vanishes, the auxiliary field equation of motion reduces to $j_{\pm} = - v_{\pm}$. In this case, the Lagrangian (\ref{integrable_family_auxiliaries}) reduces to the PCM Lagrangian (\ref{PCM_lagrangian}), and the quantity $\mathfrak{J}_\alpha$ simply becomes $j_\alpha$, which is indeed conserved in the PCM.

More generally, when the auxiliary field $v_\alpha$ has been eliminated using its equation of motion, the quantity $\fJ_{\alpha}$ becomes the Noether current associated with right-multiplication of the group-valued field $g$ by an arbitrary element of $G$. In the theory without auxiliary fields, conservation of this Noether current is equivalent to the equation of motion for the model. Just as the $4d$ IZ formalism solves the problem of parameterizing a general theory of electrodynamics which enjoys duality invariance, the family (\ref{integrable_family_auxiliaries}) corresponds (after integrating out auxiliary fields) to the most general Lagrangian depending only on $\mathrm{tr} ( j_+ j_- )$ and $\mathrm{tr} ( j_+ j_+ ) \mathrm{tr} ( j_- j_- )$ which is classically integrable with a Lax connection that takes a prescribed form (\ref{lax_connection}) in terms of this Noether current $\fJ_{\pm}$.

\section{Lax Representation}\label{sec:weak}

We say that a model admits a Lax representation if its equations of motion are equivalent to the condition
\begin{align}
    d_\fL \fL ( z ) = 0 \, , \qquad \forall z \in \mathbb{C} \, ,
\end{align}
for a Lax connection $\mathfrak{L} ( z )$ which depends meromorphically on a spectral parameter $z$. Here $d_\fL = d + \fL \, \wedge \, $ is the exterior covariant derivative associated with $\fL$. In light-cone coordinates, $d_\fL \fL$ can be expressed as
\begin{align}\label{curvature_light_cone}
    d_\fL \fL = \partial_+ \mathfrak{L}_- - \partial_- \mathfrak{L}_+ + [ \mathfrak{L}_+ , \mathfrak{L}_- ] \, .
\end{align}
Let us first define what we mean by a Lax representation for an auxiliary field model, where one has both the $v_\alpha$-field equation of motion (\ref{aux_eom}) and the $g$-field equation of motion (\ref{j_eom}). Given a Lagrangian $\mathcal{L} ( j_\alpha, v_\alpha )$, not to be confused with the Lax $\fL ( j_\alpha, v_\alpha )$, one could eliminate the auxiliary field $v_\alpha$ using its equation of motion to write
\begin{align}
    \mathcal{L} ( j_\alpha , v_\alpha ) \deq \mathcal{L} ( j_\alpha ) \, .
\end{align}
Likewise, one could eliminate the auxiliary fields in an expression for a Lax connection $\mathfrak{L} ( j_\alpha , v_\alpha )$ to obtain
\begin{align}
    \fL ( j_\alpha, v_\alpha ) \deq \fL ( j_\alpha ) \, .
\end{align}
If the flatness of the resulting Lax connection $\fL ( j_\alpha )$ for any $z \in \mathbb{C}$ is equivalent to the $g$-field equation of motion arising from $\mathcal{L} ( j_\alpha )$, we say that the original auxiliary field model $\mathcal{L} ( j_\alpha , v_\alpha )$ admits a Lax representation with Lax connection $\mathfrak{L} ( j_\alpha , v_\alpha )$. An equivalent, but more succinct, version of this definition is as follows. If we define
\begin{align}
    \mathcal{E}_g = \partial_\alpha \left( \frac{\partial \mathcal{L}}{\partial ( \partial_\alpha g ) } \right)  - \frac{\partial \mathcal{L}}{\partial g} \, ,
\end{align}
so that $\mathcal{E}_g = 0$ is the $g$-field equation of motion, then the existence of a Lax representation for an auxiliary field model is the statement
\begin{align}\label{weak_int_defn}
    \left( d_\fL \fL \deq 0 \, , \, \forall z \in \mathbb{C} \, \right) \, \iff \, \left(  \mathcal{E}_g \deq 0 \right) \, .
\end{align}
We will now demonstrate that every Lagrangian of the form (\ref{integrable_family_auxiliaries}) satisfies the condition (\ref{weak_int_defn}), where the Lax is
\begin{align}\label{lax_connection}
    \mathfrak{L}_{\pm} = \frac{j_{\pm} \pm z \mathfrak{J}_{\pm}}{1 - z^2} \, .
\end{align}
First note that the auxiliary field equation (\ref{aux_eom}) implies
\begin{align}\label{nice_commutators}
    [ \mathfrak{J}_+ , \mathfrak{J}_- ] \deq [ j_+, j_- ] \, , \qquad [ \mathfrak{J}_+ , j_- ] \deq [ j_+ , \mathfrak{J}_- ] \, ,
\end{align}
which agree with relations that were found in the study of the root-$\TT$-deformed PCM without auxiliary fields \cite{Borsato:2022tmu}. If we had also allowed the interaction function $E$ to depend on the combination $\mathrm{tr} ( v_+ v_- )$, the first relation of (\ref{nice_commutators}) would not hold and this argument would be spoiled. However, dependence on the variables $\mathrm{tr} ( v_+^k ) \mathrm{tr} ( v_-^k )$for $k > 2$ can be accommodated, as studied in \cite{Bielli:2024ach}. The formulas (\ref{nice_commutators}) allow us to simplify the commutator
\begin{align}
    [ \fL_+ , \fL_-] &= \frac{[ j_+, j_- ] - z \left( [ j_+, \fJ_- ] - [ \fJ_+ , j_- ] \right) - z^2 [ \fJ_+ , \fJ_- ] }{ \left( 1 - z^2 \right)^2} \nonumber \\
    &\deq \, \frac{ [j_+ , j_-] }{1 - z^2} \, . 
\end{align}
The curvature (\ref{curvature_light_cone}) of the Lax connection (\ref{lax_connection}) is then
\begin{align}\label{lax_curvature}
    d_\fL \fL \deq \frac{\partial_+ j_- - \partial_- j_+ + [ j_+, j_-] - z \left( \partial_+ \fJ_- + \partial_- \fJ_+ \right)}{1 - z^2} \, .
\end{align}
The first three terms in the numerator of (\ref{lax_curvature}) vanish due to the Maurer-Cartan identity (\ref{maurer_cartan}), and the final two terms vanish if and only if $\partial_\alpha \fJ^\alpha \deq 0$, which is the equation of motion for $g$. We conclude that every model in this family obeys the condition (\ref{weak_int_defn}).

\section{Classical Integrability}\label{sec:strong}

We say that a field theory is classically integrable if the theory possesses an infinite set of conserved charges which are in involution, or mutually Poisson-commuting. Demonstrating that the equations of motion for a model admit a Lax representation is a useful step in this direction, since one can then construct an infinite set of charges using the monodromy matrix. However, the Poisson-commutativity of these charges is not guaranteed unless the Poisson bracket of the Lax connection,
\begin{align}\label{general_lax_lax_bracket}
    \left\{ \fL_{\sigma, 1} ( \sigma, z ) , \fL_{\sigma, 2} ( \sigma' , z' ) \right\} \, ,
\end{align}
takes a special form. Here we have introduced, for any Lie algebra valued quantity $X$, the subscript notation
\begin{align}
    X_1 = X \otimes 1 \, , \qquad X_2 = 1 \otimes X \, ,
\end{align}
which tensors $X$ with the identity on either side. We do not carefully distinguish between $\mathfrak{g}$ and $U ( \mathfrak{g} )$, the universal enveloping algebra of $\mathfrak{g}$, but strictly speaking $1 \in U ( \mathfrak{g} )$ and $X_1, X_2 \in U ( \mathfrak{g} ) \otimes U ( \mathfrak{g} )$.
If (\ref{general_lax_lax_bracket}) takes a Sklyanin \cite{Sklyanin:1980ij} or Maillet \cite{MAILLET198654,MAILLET1986401} form, then it follows that an infinite subset of the conserved charges constructed from the monodromy matrix are in involution.

We will now prove that, for every model in the family (\ref{integrable_family_auxiliaries}), the Poisson bracket (\ref{general_lax_lax_bracket}) takes the non-ultralocal Maillet form. This argument proceeds almost exactly as in the analogous proof for the PCM \cite{MAILLET1986401}, except replacing every instance of $j_\tau$ with $\fJ_\tau$; see \cite{Driezen:2021cpd} for a review.

To study the canonical structure of models with group-valued fields \cite{BOWCOCK198980}, it is convenient to introduce local coordinates $\phi^\mu$ on the Lie group $G$, not to be confused with $\varphi(z)$, so that $g = g ( \phi^\mu ( \sigma^\alpha ) )$. We use early Greek letters like $\alpha$, $\beta$ for indices on $\Sigma$ and middle Greek letters like $\mu$, $\nu$ for coordinates on $G$. We also introduce capital early Latin letters (e.g. $A$, $B$) which label the generators $T_A$ of the Lie algebra $\mathfrak{g}$. For any Lie algebra valued quantity $X$, one can expand in generators as $X = X^A T_A$. Furthermore, we may use the pull-back map $\partial_\alpha \phi^\mu$ to convert between indices on $\Sigma$ and indices on $G$. For instance,
\begin{align}
    j_\alpha = j_\mu^A \frac{\partial \phi^\mu}{\partial \sigma^\alpha} T_A \, .
\end{align}
We write $\gamma_{AB}$ for the Killing form on $\mathfrak{g}$, which we assume to be non-degenerate with inverse $\gamma^{AB}$, and we denote the structure constants by $\tensor{f}{_A_B^C}$. In our conventions, these two objects are defined by the relations
\begin{align}
    \gamma_{AB} = \Tr{ T_A T_B } \, , \qquad  [ T_A, T_B ] = \tensor{f}{_A_B^C} T_C \, .
\end{align}
In terms of these quantities, the canonical momentum $\pi_\mu$ which is conjugate to the coordinates $\phi^\mu$ on $G$ is
\begin{align}
    \pi_\mu = \frac{\partial \mathcal{L}}{\partial ( \partial_\tau \phi^\mu ) } = j_\mu^A \left( j_\tau^B + 2 v_\tau^B \right) \gamma_{AB} \, .
\end{align}
This means that the quantity $\fJ_\tau^A = - ( j_\tau^A + 2 v_\tau^A )$ is simply related to the canonical momentum as
\begin{align}\label{J_to_pi}
    \fJ_\tau^A = - \gamma^{AB} \pi_\mu j^\mu_B \, ,
\end{align}
where we have defined the inverse field $j^\mu_B$ which satisfies $j^A_\mu j^\mu_B = \tensor{\delta}{^A_B}$. The fundamental Poisson brackets are then
\begin{align}\label{ccr}
    \left\{ \pi_\mu ( \sigma ) , \phi^\nu ( \sigma' ) \right\} = \tensor{\delta}{_\mu^\nu} \delta ( \sigma - \sigma' ) \, ,
\end{align}
which hold at equal times $\tau$. Because no time derivatives of the auxiliary field $v_\alpha$ appear in the Lagrangian, the momentum $\mathfrak{p}^\alpha$ which is conjugate to $v_\alpha$ vanishes:
\begin{align}\label{pi_v_constraint}
    \mathfrak{p}^\alpha = \frac{\partial \mathcal{L}}{\partial ( \partial_\tau v_\alpha ) } = 0 \, .
\end{align}
In the canonical formulation, our model (\ref{integrable_family_auxiliaries}) is therefore a constrained Hamiltonian system, for which (\ref{pi_v_constraint}) is a primary constraint. This structure is reminiscent of the Hamiltonian formulation of Maxwell theory, where the temporal gauge field $A^0$ also has vanishing conjugate momentum. A systematic treatment of the Hamiltonian structure of our models would proceed using the method of Dirac \cite{Dirac:1964:LQM}, which implements the primary constraint in the Hamiltonian via a Lagrange multiplier, and then imposes that the primary constraint is preserved under time evolution. However, we will not undertake a detailed discussion of this procedure here, since it is not necessary for our present purposes. Because we are only interested in computing (Poisson or Dirac) brackets involving $j_\alpha$ and $\fJ_\alpha$, where the latter is related to the canonical momentum by (\ref{J_to_pi}), it suffices to compute these brackets using the fundamental relations (\ref{ccr}) and ignore the fact that $\fJ_\alpha$ is itself composed from a constrained auxiliary field. In fact, the auxiliary field equation of motion will play no role whatsoever in the following calculations.

Using the fundamental brackets (\ref{ccr}), one can show
\begin{align}\label{nice_brackets}
    \{ \fJ_\tau^A ( \sigma ) , \fJ_\tau^B ( \sigma' ) \} &= \tensor{f}{^A^B_C} \fJ_\tau^C \delta ( \sigma - \sigma' ) \, , \nonumber \\
    \left\{ \fJ_\tau^A ( \sigma ) , j_\sigma^B ( \sigma' ) \right\} &= \tensor{f}{^A^B_C} j_\sigma^C ( \sigma ) \delta ( \sigma - \sigma' ) - \gamma^{AB} \delta' ( \sigma - \sigma' ) \, ,  \nonumber \\
    \left\{ j_\sigma^A ( \sigma ) , j_\sigma^B ( \sigma' ) \right\} &= 0 \, ,
\end{align}
where $\delta' ( \sigma - \sigma' ) = \partial^{(\sigma)} \delta ( \sigma - \sigma' )$. To obtain these results, we have used the Maurer-Cartan identity (\ref{maurer_cartan}) in the form
\begin{align}\label{mc_push_forward}
    \partial_\mu j_\nu^A - \partial_\nu j_\mu^A + j_\mu^B j_\nu^C \tensor{f}{_B_C^A} = 0 \, ,
\end{align}
along with $\delta$-function identities such as
\begin{align}
    f ( y ) \partial^{(x)} \delta ( x - y ) &= f ( x ) \partial^{(x)} \delta ( x - y ) + f' ( x ) \delta ( x - y ) \, .
\end{align}
One can then contract each of the relations (\ref{nice_brackets}) with the tensor product $T_A \otimes T_B$ to find
\begin{align}\label{Jt_js_contracted}
    \{ \fJ_{\tau,1} ( \sigma ) , \fJ_{\tau,2} ( \sigma' ) \} &= [ \fJ_{\tau, 2} , C_{12} ]  \delta ( \sigma - \sigma' ) \, , \nonumber \\
    \left\{ \fJ_{\tau, 1} ( \sigma ) , j_{\sigma, 2} ( \sigma' ) \right\} &= [ j_{\sigma, 2}, C_{12} ] \delta ( \sigma - \sigma' ) - C_{12} \delta' ( \sigma - \sigma' ) \, , \nonumber \\
    \left\{ j_{\sigma, 1} ( \sigma ) , j_{\sigma, 2} ( \sigma' ) \right\} &= 0 \, ,
\end{align}
where we have defined the Casimir $C_{12} = \gamma^{AB} T_A \otimes T_B$.

From (\ref{Jt_js_contracted}) one can show that
\begin{widetext}
\begin{align}\label{final_lax_bracket}
    \left\{ \fL_{\sigma, 1} ( \sigma, z ) , \fL_{\sigma, 2} (\sigma', z' ) \right\} &= \frac{1}{( 1 - z^2 ) ( 1 - z^{\prime 2} ) } \Big( \left[ z z' \fJ_{\tau, 2}  + ( z + z' ) j_{\sigma, 2 } , C_{12} \right] \delta ( \sigma - \sigma' ) - ( z + z' ) C_{12} \delta' ( \sigma - \sigma' ) \Big) \nonumber \\
    &= \left[ r_{12} ( z, z' ) , \fL_{\sigma, 1} ( \sigma, z ) \right] \delta ( \sigma - \sigma' )  - \left[ r_{21} ( z', z ) , \fL_{\sigma, 2} ( \sigma, z' ) \right] \delta ( \sigma - \sigma' ) - s_{12} ( z, z' ) \delta' ( \sigma - \sigma' ) \, ,
\end{align}
\end{widetext}
where $s_{12} ( z , z' ) = r_{12} ( z, z' ) + r_{21} ( z', z )$, the $r$-matrix is
%
\begin{align}
    r_{12} ( z, z' ) &= \frac{C_{12}}{z - z'} \varphi^{-1} ( z' ) \, , 
\end{align}
which solves the classical Yang-Baxter equation, and
\begin{align}
    \varphi ( z ) &= \frac{ z^2 - 1 }{z^2} 
\end{align}
is the same twist function as for the undeformed PCM. The structure in the second line of (\ref{final_lax_bracket}) is precisely the desired non-ultralocal Maillet form of the Poisson bracket. 

One may now construct an infinite set of Poisson-commuting conserved charges in exactly the same way as is done for the PCM, which we sketch only briefly. One first defines a transport matrix via a path-ordered exponential of the spatial Lax connection. Taylor expanding a limit of this transport matrix around an appropriate value of $z$ gives rise to an infinite set of conserved charges, whose Poisson brackets are related to those of the Lax connection. When the Poisson bracket of the Lax takes the form (\ref{final_lax_bracket}), by regulating the coincident-point limit of the transport matrices using a symmetrization procedure described in \cite{MAILLET198654}, one can show that an infinite subalgebra of these conserved charges are in involution \footnote{See equation (3.24) of \cite{MAILLET198654}, and the preceding discussion, for an explicit expression for these conserved charges.}. Carrying out this procedure for the models (\ref{integrable_family_auxiliaries}) establishes the existence of an infinite collection of independent Poisson-commuting conserved charges, for any choice of interaction function, so we conclude that every model in this family is classically integrable.

\section{Stress Tensor Flows}\label{sec:stress}

The family of sigma models (\ref{integrable_family_auxiliaries}) is classically equivalent to the set of all deformations of the PCM by functions of the stress tensor, which we define by
\begin{align}
    T_{\alpha \beta} = - \frac{2}{\sqrt{-g}} \frac{\delta S}{\delta g^{\alpha \beta}} \, .
\end{align}
In components, $\tensor{T}{^\alpha_\beta}$ is a $2 \times 2$ matrix, which means that there are two functionally independent Lorentz scalars that can be constructed from it. One is the trace,
\begin{align}
    \tensor{T}{^\alpha_\alpha} \deq 2 \left( E - 2 E' \nu \right) \, ,
\end{align}
where we have simplified using the auxiliary field equation of motion (\ref{aux_eom}). The other is $\tensor{T}{^\alpha_\beta} \tensor{T}{^\beta_\alpha}$, which satisfies
\begin{align}
    \tensor{T}{^\alpha_\beta} \tensor{T}{^\beta_\alpha} - \frac{1}{2} \left( \tensor{T}{^\alpha_\alpha} \right)^2 \deq \frac{1}{2} \nu \left( 1 - 4 ( E' )^2 \nu \right)^2 \, .
\end{align}
Importantly, both of these contractions can be expressed entirely in terms of $\nu$, $E ( \nu )$, and $E' ( \nu )$, when the auxiliary field equation of motion is satisfied. This means that any classical flow equation of the form
\begin{align}\label{classical_flow}
    \frac{\partial \mathcal{L}}{\partial \lambda} \deq f \left( \tensor{T}{^\alpha_\alpha} , \tensor{T}{^\alpha_\beta} \tensor{T}{^\beta_\alpha} , \lambda \right) \, ,
\end{align}
gives rise to an ordinary differential equation for the interaction function $E ( \nu )$, whose solution is another model within the same class of theories. This structure is similar to that of classical stress tensor deformations of theories of duality-invariant electrodynamics in $4d$ \cite{Ferko:2023wyi}, chiral tensor theories in $6d$ \cite{Ferko:2024zth}, or chiral bosons in $2d$ \cite{Ebert:2024zwv}.

Examples of deformations (\ref{classical_flow}) include the $\TT$ \cite{Zamolodchikov:2004ce,Cavaglia:2016oda} and root-$\TT$ \cite{Ferko:2022cix,Babaei-Aghbolagh:2022leo,Babaei-Aghbolagh:2022uij} flows, driven by the functions
\begin{align}\label{TT_flow}
    \mathcal{O}_{\TT} &= \tensor{T}{^\alpha_\beta} \tensor{T}{^\beta_\alpha} - \left( \tensor{T}{^\alpha_\alpha} \right)^2 \, , \\
    \label{root_TT_flow} \mathcal{R} &= \frac{1}{\sqrt{2}} \sqrt{ \tensor{T}{^\alpha_\beta} \tensor{T}{^\beta_\alpha} - \frac{1}{2} \left( \tensor{T}{^\alpha_\alpha} \right)^2 } \, .
\end{align}
The solution for the interaction function $E ( \nu )$ deformed by the root-$\TT$ flow driven by $\mathcal{R}$, subject to the initial condition $E ( \lambda = 0 , \nu ) = 0$, is
\begin{align}\label{modmax_E}
    E ( \lambda, \nu ) = \tanh \left( \frac{\lambda}{2} \right) \sqrt{ \nu } \, ,
\end{align}
which is the same as the interaction function for the ModMax theory \cite{Bandos:2020jsw} in the $4d$ Ivanov-Zupnik formulation \cite{Kuzenko:2021cvx}. We note that an interaction function proportional to $\sqrt{\nu}$, such as (\ref{modmax_E}), is the only choice compatible with classical conformal invariance. The solution to the flow equation driven by the operator $\mathcal{O}_{\TT}$ of (\ref{TT_flow}), with the same initial condition, can be written in terms of a hypergeometric function \cite{Ferko:2024zth,Bielli:2024ach}. A third example is a linear interaction function $E ( \nu ) \sim \nu$, which is a $2d$ analogue of the ``simplest interaction model'' (or Bossard-Nicolai model) that has been studied in $4d$ electrodynamics \cite{Ivanov:2012bq,Bossard:2011ij,Carrasco:2011jv}.

It is well-known that applying the $\TT$ deformation to a classically integrable theory produces another classically integrable theory \cite{Smirnov:2016lqw}, and the Lax connection for the $\TT$-deformed PCM was obtained in \cite{Chen:2021aid}. Likewise, the Lax connection for the PCM deformed by both $\TT$ and root-$\TT$ appeared in \cite{Borsato:2022tmu}. Both of these results can be viewed as special cases of the general Lax connection (\ref{lax_connection}), which is appropriate for a deformation of the PCM by an arbitrary function of the energy-momentum tensor.

\vspace{10pt}

\section{Conclusions and Outlook}\label{sec:conclusion}

In this work, we introduced a new auxiliary field technique for proving classical integrability of $2d$ sigma models. We used this machinery to establish integrability for an infinite class of PCM-like theories, which includes all deformations of the PCM by functions of the energy-momentum tensor. By virtue of the auxiliary fields in our model, the proof of integrability for this entire infinite class is no more difficult than that for the PCM.

Our discussion represents a case study of a much more general lesson in field theory, which applies beyond the setting of integrability. The take-away message is that, by coupling a simpler theory to auxiliary fields as in (\ref{integrable_family_auxiliaries}), one can often obtain a larger class of deformed theories which preserve some desirable structure of the undeformed model. It was already known that such a prescription can introduce interactions in theories of $4d$ electrodynamics while preserving duality invariance \cite{Ivanov:2002ab,Ivanov:2003uj}, or in $6d$ tensor theories while preserving chirality \cite{Ferko:2024zth}.

We have illustrated that a similar technique can be applied in the seemingly-unrelated context of $2d$ integrable theories. Our strategy has also been applied \cite{Fukushima:2024nxm} to $4d$ Chern-Simons theory \cite{Costello:2019tri,nekrasov_thesis,Costello:2013zra,Lacroix:2021iit}, whose deformations were previously connected to $\TT$ \cite{Py:2022hoa}, and to the study of non-Abelian T-duality \cite{Bielli:2024khq}. We believe that this approach will be useful in other domains. Analogous auxiliary field couplings might be used to generate families of deformed boundary conditions in holography, or to deform QFTs in a way which preserves generalized global symmetries. We hope that the philosophy advocated in this work will inspire the use of related techniques in other field theories which find applications in classical and quantum gravity, condensed matter physics, and phenomenology.

\begin{acknowledgments}

We thank Gabriele Tartaglino-Mazzucchelli and Alessandro Sfondrini for helpful discussions and comments on a draft of this letter. We also acknowledge kind hospitality and financial support at the MATRIX Program ``New Deformations of Quantum Field and Gravity Theories'' and thank the participants of this meeting for productive conversations. C.\,F. is supported by U.S. Department of Energy grant DE-SC0009999 and funds from the University of California. L.\,S. is supported by a postgraduate
scholarship at the University of Queensland and by Australian Research Council (ARC) grants  FT180100353 and DP240101409. This research was supported in part by grant NSF PHY-2309135 to the Kavli Institute for Theoretical Physics (KITP).

\end{acknowledgments}

\appendix

\bibliographystyle{apsrev4-2} 
\bibliography{apssamp}

\end{document}